\documentclass[pdflatex,sn-nature]{sn-jnl}
\usepackage{graphicx}%
\usepackage{multirow}%
\usepackage{amsmath,amssymb,amsfonts}%
\usepackage{amsthm}%
\usepackage{mathrsfs}%
\usepackage[title]{appendix}%
\usepackage{xcolor}%
\usepackage{textcomp}%Nat
\usepackage{manyfoot}%
\usepackage{booktabs}%
\usepackage{algorithm}%
\usepackage{algorithmicx}%
\usepackage{algpseudocode}%
\usepackage{listings}%
\newcommand{\mat}[1]{\ensuremath{\boldsymbol{\mathrm{#1}}}}
\renewcommand{\d}{\mathrm{d}}
\newcommand{\sigman}{\sigma_{\rm n}}
\theoremstyle{thmstyleone}%
\theoremstyle{thmstyletwo}%
\theoremstyle{thmstylethree}%
\begin{document}
%\title{The prominent role of seepage forces in creeping landslides: Example from \AA knes, Norway}
\title{Unveiling the role of seepage forces in the acceleration of landslides creep}
%\title{How seepage forces can govern creep acceleration in landslides}
%
\author*[1]{\fnm{Fabian} \sur{Barras}}\email{fabian.barras@mn.uio.no}
\equalcont{These authors contributed equally to this work.}
\author*[1,2]{\fnm{Andreas} \sur{Aspaas}}\email{agas@nve.no}
\equalcont{These authors contributed equally to this work.}
\author[1,3]{\fnm{Einat} \sur{Aharonov}}\email{einatah@mail.huji.ac.il}
\author[1,4]{\fnm{Fran\c{c}ois} \sur{Renard}}\email{francois.renard@mn.uio.no}
\affil[1]{\orgdiv{The Njord Centre, Departments of Geosciences and Physics}, \orgname{University of Oslo}, \postcode{0316} \city{Oslo}, \country{Norway}}
\affil[2]{\orgdiv{Section for Landslides and Avalanches}, \orgname{Norwegian Resources and Energy Directorate}, \postcode{0368} \city{Oslo}, \country{Norway}}
\affil[3]{\orgdiv{Institute of Earth Sciences}, \orgname{The Hebrew University}, \postcode{91904} \city{Jerusalem}, \country{Israel}}
\affil[4]{\orgdiv{ISTerre}, \orgname{Univ. Grenoble Alpes, Grenoble INP, Univ. Savoie Mont Blanc, CNRS, IRD, Univ. Gustave Eiffel}, \postcode{38000} \city{Grenoble}, \country{France}}
%
%%==================================%%
%% Sample for unstructured abstract %%
%%==================================%%
%
%%================================%%
%% Sample for structured abstract %%
%%================================%%
% Limit: 200 words
\abstract{In the context of global climate change, geological materials are increasingly destabilized by water flow and infiltration. We study the creeping dynamics of a densely monitored landslide in Western Norway to decipher the role of fluid flow in destabilizing this landslide. In \AA knes, approximately 50 million cubic meter of rock mass continuously creeps over a shear zone made of rock fragments, with seasonal accelerations that strongly correlate with rainfall. In this natural laboratory for fluid-induced frictional creep, unprecedented monitoring equipment reveals low fluid pressure across the shear zone, thereby challenging the dominant theory of fluid-driven instability in landslides. Here, we show that a generic micromechanical model can disentangle the effects of fluid flow from those of fluid pressure, and demonstrate that seepage forces applied by channelized flow along the shear zone are the main driver of creep accelerations. We conclude by discussing the significance of seepage forces, the implications for hazard mitigation and the broader applicability of our model to various geological contexts governed by friction across saturated shear zones.}
\keywords{Landslide, Creep burst, Friction, Pore pressure, Seepage, Shear zone rheology}
\maketitle
The stability of natural slopes is a critical concern in the context of climate change, as precipitations and fluid infiltration increasingly perturb landslides worldwide. The zones of weakness along which slopes deform are typically made of unconsolidated debris and clay-rich material. Often these granular shear zones do not stand still and commonly deform at a slow creep rate, in the range of millimetres to centimetres per year \cite{veveakis2007thermoporomechanics,groneng2009shear,marcato2012monitoring,walter2013slidequake,roman2020creeping}. This slow and mostly aseismic deformation may be punctuated by short-lived accelerations called creep bursts. These bursts are often driven by intense rainfall or snow melt, accumulating strain that may precede runaway failure \cite{cappa2004hydromechanical,alonso2015thermo,tang2019geohazards,krishnendu2024frictional}. Understanding how fluids interact with the granular shear zones of landslides is therefore essential for linking rainfall and snow-melt to landslide dynamics and for hazard assessment and  mitigation.

For decades, the dominant framework has been Terzaghi’s effective-stress theory, in which elevated pore fluid pressure reduces the normal stress and thereby weakens the frictional resistance of soils and rocks \cite{terzaghi_theoretical_1943}. The effect of pore fluid pressurization is expected to play a major role in both the initiation and acceleration preceding failure of landslides \cite{iverson1987rainfall,iverson1986groundwater,goren2007long,goren2009stability,Viesca2012nucleation}. Yet, in the context of creeping landslides, recent in-situ measurements reveal how rainfall-induced creep bursts arise with negligible fluid pressure in the shear zones \cite{aspaas2024}, 
suggesting that pore pressure alone cannot always account for the observed seasonal creep bursts. This discrepancy motivates consideration of an additional hydro-mechanical effect, that of seepage force exerted by water flow within a granular shear zone.

\begin{figure}[!ht]
\centering
\includegraphics[width=0.95\linewidth]{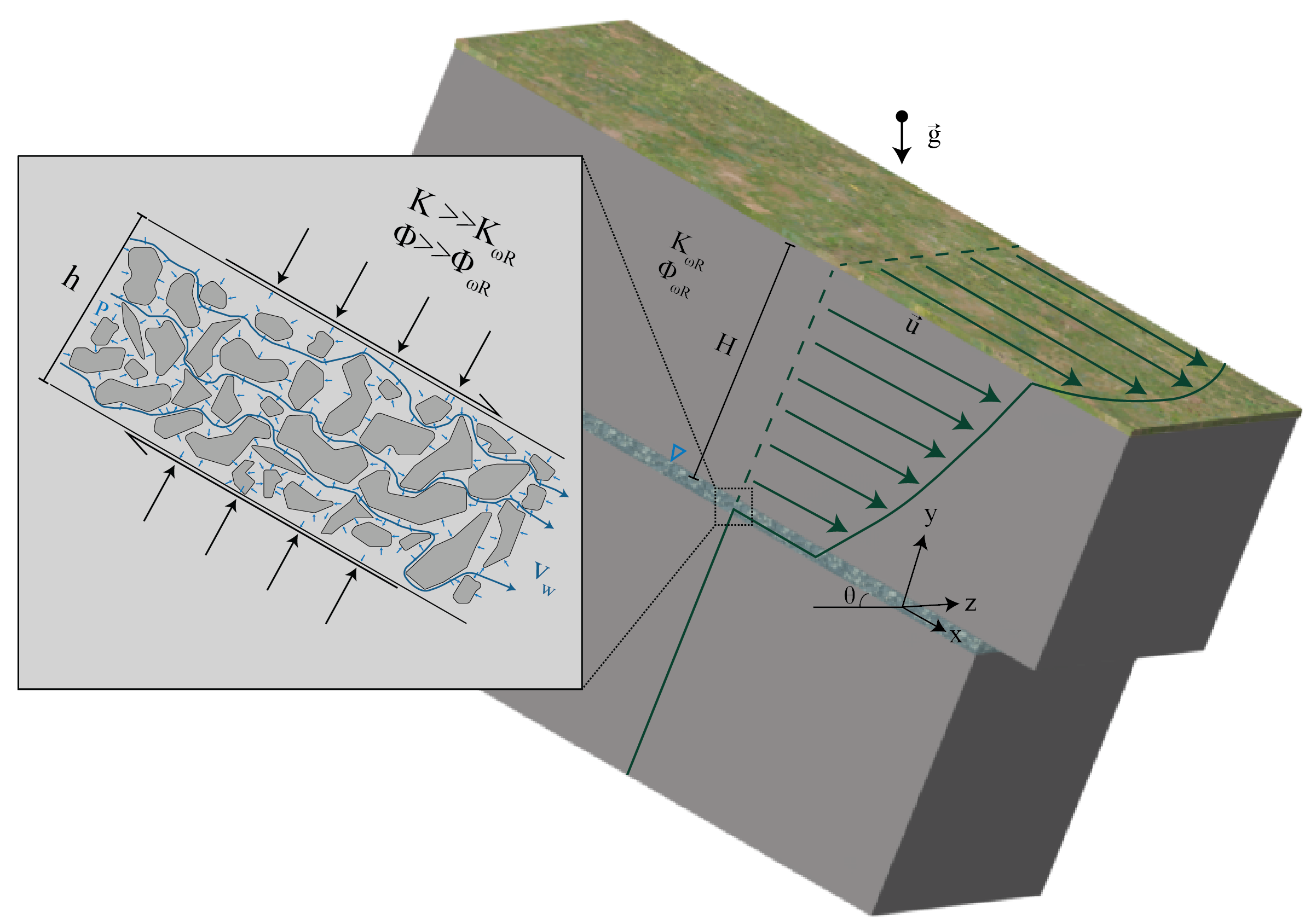}
\caption{Geometry of a typical creeping landslide studied in this work. A rock mass of thickness $H$ slides downhill following a slip profile $u$ where most of the strain is localised within a shear zone of thickness $h$ made up of debris and rock particles crushed by frictional sliding. Due to its granular nature, the damaged shear zone has a larger porosity $\phi$ and hydraulic conductivity $K$ than the surrounding competent rock mass ($K_{wr}$,$\phi_{wr}$) and represents a favourable flow corridor for underground water. The inset details the effect of water at the grain scale of the shear zone, which is twofold: 1) a hydrostatic effect caused by the pore pressure $p$, which reduces the effective normal stress and 2) the seepage force exerted by fluid flowing at an average rate $v_{\rm w}$ on the solid particles, which increases the shear stress at the base of the landslide.}
\label{fig:concept}
\end{figure}

To disentangle and clarify the full spectrum of hydro-mechanical forces, we develop a generic mechanical model, combining basic stress balance with granular and fluid flow in a creeping granular shear zone, Fig \ref{fig:concept}. The model separates hydrostatic pressurization from flow-induced seepage forces and tests their respective impacts on landslide creep rates. To evaluate the model, we apply it to a creeping landslide in Norway. This natural laboratory offers a unique opportunity to test our approach due to three key features: (1) displacement is localized on shear zones comprising debris and clay-rich material; (2) the landslide exhibits seasonal creep bursts that correlate with intense rainfall events; and (3) deformation is monitored using state-of-the-art high-resolution instrumentation mounted in an array of boreholes. Despite low fluid pressures, visual evidence indicates significant fluid flow through the shear zones, supporting the relevance of seepage-driven dynamics (see video in Supplemental Material).

The forces caused by seepage through soil is well-established in engineering and is accounted in the design of geotechnical infrastructures, such as embankment dams and retaining walls. Seepage exerts a similar forcing in the deformation of landslides if much of the moving body is strained by macroscopic fluid flow \cite{cappa2004hydromechanical,tacher2005modelling,tang2019geohazards}. However, the micromechanical effect of seepage and its coupling to the rheology, the stability and the creep dynamics of localized shear zone remains overlooked. In this work, we address the need for a generic analytical tool, describing the combined effects of seepage and fluid pore-pressure, in driving water-assisted landslide creep. This model may be used to predict landslide motion and improving landslide hazard mitigation. The predictive capacity of the proposed model is demonstrated by matching creep field data of the \AA knes landslide in Norway and establishes how seepage forces operating only within the thin shear zone of a landslide can nonetheless exert a major control on seasonal creep bursts.
\section*{Modelling shear zone creep including granular flow and seepage forces}
We construct a generic mechanical model of creeping landslide. A rock mass of thickness $H$ creeps down an inclined plane, with most of the shear strain localised into a shear zone of thickness $h$, leading to the profile of creep $\vec{u}=u_x\vec{e}_x$ sketched in Figure \ref{fig:concept}. The deformation of the system is two-fold, since the creep rate $\dot{u}_x=\partial u_x/\partial t$ results from the sum of elastic deformation of the rock mass and viscoplastic deformation localised in the shear zone comprising unconsolidated debris (e.g., Fig \ref{fig:setup}c). Different approaches have been proposed in the literature to model the rheology of a granular shear zone, such as (1) viscoplastic models related to the Mohr-Coulomb description \cite{tacher2005modelling,veveakis2007thermoporomechanics}, (2) friction models derived from a rate-and-state constitutive law \cite{agliardi2020slow}, (3) granular flow model deriving from the $\mu(I)$-rheology \cite{jop_constitutive_nature_2006,parez2021strain,Blatny2024critical}, and (4) Bingham fluid model used to describe the rheology of clay \cite{locat1988viscosity}. These approaches share a Mohr-Columb yield criterion, which linearly depends on the normal stress $\sigman$ minus the pore pressure $p$, combined with a flow rule beyond the yield point. In this context, we postulate the following generic constitutive model for the rheology of the shear zone:
\begin{equation}
\left\lbrace
\begin{array}{c}
\tau \leq (\sigman-p)\tan\psi, \;\;\mathrm{for}\; \dot{\gamma}=0\\
\tau = (\sigman-p)\tan\psi + \mu\dot{\gamma}, \;\;\mathrm{for}\; |\dot{\gamma}|>0.
\label{equ:viscous_model}
\end{array}
\right.
\end{equation}
Two parameters define the friction constitutive law (\ref{equ:viscous_model}): a cohesion-less yield stress defined by the angle of repose $\psi$ of the unconsolidated shear zone, and a shear-zone granular effective viscosity that produces shear stress $\tau$ from the shear strain rate $\dot{\gamma}=\partial \dot{u}_x/\partial y$ beyond the yield stress. For the mild variations in creep strain rate studied in this paper, the viscous model (Eq. \ref{equ:viscous_model}) can be understood as a linearized approximation of the rheology of the shear zone.

In the Methods, we detail how poroelasticity coupled to the permeable shear zone rheology (Eq. \ref{equ:viscous_model}) and Darcy flow through it (Eq. \ref{equ:darcy}), can be used to express the stress balance governing the creep dynamics of the system (Eq. \ref{equ:x-momentum}). Under the assumptions that the thickness of the rock mass $H$ is much smaller than the other dimensions of the landslide, the stress balance is solved on average across $H$ to obtain the following \textit{creep equation} of the sliding body, written in terms of the average creep rate $\dot{u}$ and displacement $u$ (see Eq. \ref{equ:average_displ}):
\begin{equation}
      \eta\dot{u} = \tan\theta(1 - \alpha + \alpha \lambda) + \kappa\Big(\frac{2}{1-\nu}\frac{\partial^2 u}{\partial x^2}+\frac{\partial^2 u}{\partial z^2}\Big)+ \omega v_{\rm w} .
    \label{equ:creep_equ}
\end{equation}
Equation (\ref{equ:creep_equ}) is fully derived, including details regarding the prefactors, as Eq. (\ref{equ:creep-rate-full}) in Methods.
On the left hand side of Eq. (\ref{equ:creep_equ}), the frictional stress due to velocity-strengthening shear zone creep is expressed by the viscous coefficient $\eta$, and by the creep rate $\dot{u}$. It balances the driving forces on the right hand side of Eq. (\ref{equ:creep_equ}) that result from three contributions: 1) the gravitational driving force, characterized by the slope factor of safety 
$\alpha=\tan \psi/\tan \theta$ (ratio of tangents of the angle of repose and slope angle), 2) the elastic interactions across the landslide which depend on the second-order spatial derivatives of the displacement, the elastic coefficient $\kappa$ and the Poisson's ratio $\nu$ of the rock mass, and 3) infiltrated water that can destabilize the landslides and accelerate creep via two distinct mechanisms. Fluid pressurisation weakens the shear zone by reducing effective normal stress (Eq. \ref{equ:viscous_model}) and is quantified by the dimensionless ratio $\lambda=p/\sigma_n$. In addition, pore fluid flow also induces seepage forces in the shear plane that scale as the average fluid flow rate across the shear zone, $v_{\rm w}$, multiplied by the seepage coefficient, $\omega$. The full expressions for each coefficient of the creep equation (\ref{equ:creep_equ}) are detailed in Table \ref{Tab:dimensionless_SI}, and in Methods. To demonstrate the capability of the proposed model, we apply it to study the creep dynamics of the \AA knes landslide, Norway.
\section*{Creep behaviour and correlation with water infiltration in the \AA knes landslide, Norway}
\begin{figure}[!ht]
\centering
\includegraphics[width=0.95\linewidth]{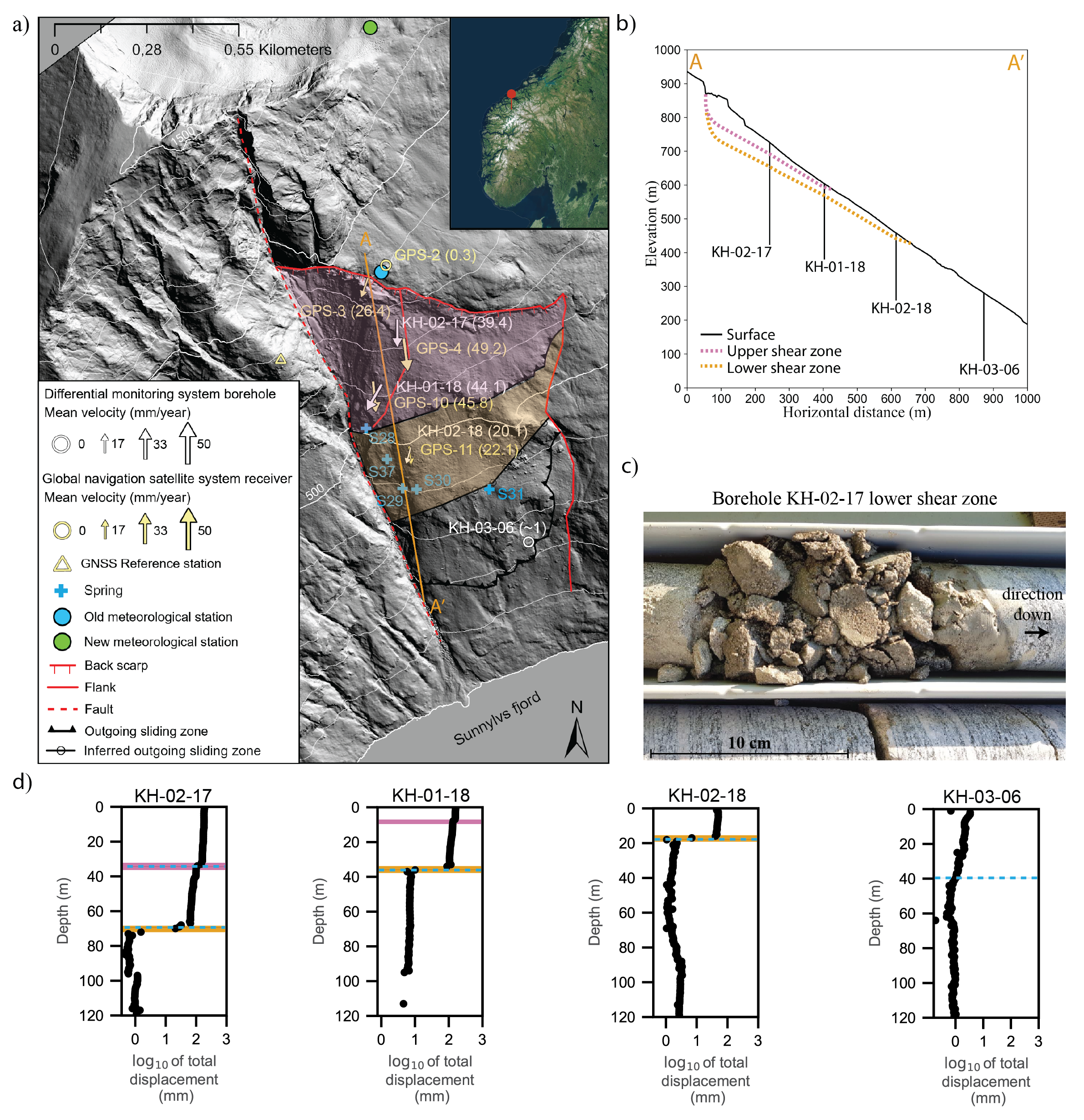}
\caption{The creeping landslide of \AA knes in Norway is closely monitored for the risk of inducing a catastrophic tsunami in the Sunnylvs fjord. (a) Morphological map of the \AA knes landslide with monitoring instruments. Borehole (white) and GNSS (yellow) arrows indicate mean velocities for the observation period (2020-2022), scaled by magnitude. Two sliding scenarios are outlined by the purple and orange polygons and are respectively associated with the upper and lower shear zones shown along the transect A–A$'$ in panel (b). (c) The structure of the upper shear zone revealed by the drill log of the borehole KH-02-17 at 69.7 m depth comprises clay-rich debris and crushed rock. (d) Cumulative slip profile (in log axis) for the period 2020-2022 was recorded at different depths of each borehole, where the upper and lower shear zones as well as the groundwater level are respectively marked by the horizontal purple, orange, and dashed blue lines. The water head coincides with the depth of the shear zones that represent favourable flow path of larger permeability.}
\label{fig:setup}
\end{figure}
On the west coast of Norway, along the UNESCO world heritage-listed Sunnylvs fjord, lies the \AA knes landslide, Norway's most renowned unstable slope. This site poses a significant hazard to fjord-side communities due to the potential for catastrophic collapse, which could generate a large tsunami \cite{pless2021possibility}. The landslide features a prominent graben-like back-scarp reaching up to 20 meters in height and 30 meters in width. Its lateral boundaries consist of strike-slip and west-dipping faults. Three sliding surfaces have been identified within the study area. Two outcropping sliding surfaces were observed directly in the field through visual inspection. One corresponds to the purple plane in Figure \ref{fig:setup}a and b, while the other is an inactive outcropping sliding plane located 200 meters above the sea (Figure \ref{fig:setup}a). A third sliding plane, represented by the yellow plane in Figure \ref{fig:setup}a and b, is inferred by geophysical survey data. The purple and yellow sliding planes are visible in the core logs as well as in the displacement data from the borehole instruments (Figure \ref{fig:setup}d). The largest potential volume to collapse includes the lowermost inactive sliding plane and is estimated to be 54 million m$^{3}$ \cite{pless2021possibility}. The landslide's bedrock is primarily made of orthogneiss, which contains locally up to 20 cm-thick biotite schist layers. Most fractures and the foliation angles are slope parallel, i.e., falling towards the fjord \cite{elvebakk2018borehullslogging}. The drill cores display clay-rich shear zones in biotite-rich gneiss \cite{langeland2019a,langeland2019b}. The structure of the lower shear zone in borehole KH-02-17 is shown in Figure \ref{fig:setup}c.

Since 2007, this landslide has been continuously monitored with high-resolution instrumentation \cite{aspaas2024}. The present investigation focuses on four boreholes (labelled KH-02-17, KH-01-18, KH-02-18, and KH-03-06 see Figure \ref{fig:setup}), which provide displacement measurements at 1-meter intervals down to a depth of 100 meters. Below this depth, measurements are recorded more sporadically, extending to total depths ranging from 105 to 190 meters. These depths extend well beyond the basal sliding surfaces, which are located between 9 and 69 meters in the boreholes. A notable feature of the monitoring setup is the absence of protective plastic tubing during borehole installation, allowing a direct mechanical connection between the instruments and the surrounding rock. Air-inflatable packers isolate fractured zones, enabling groundwater pressure sensors to measure the water pressures within shear zones (Figure 2 in \cite{aspaas2024}). All sensors operate at an hourly sampling frequency, and further details on data acquisition and processing are provided in \cite{aspaas2024}. The present investigation draws on a two-year dataset (01.01.2020 to 01.01.2022) acquired from the four boreholes. The analysis concentrates on transient accelerations within the sliding zone, aiming to elucidate the mechanisms driving these creep bursts.

Deformation within the landslide is localised along one or two shear zones, each comprising a ten-centimetre thick layer of debris, silt, and clay with negligible cohesion (Figure \ref{fig:concept} and \ref{fig:setup}). The estimated angle of repose ($\psi\in[25^\circ, 29^\circ]$) is slightly lower than the average terrain slope ($\theta\approx30^\circ$), consistent with continuous creep at a rate of a few centimetres per year \cite{langeland2014utvikling,groneng2009shear}. Episodes of accelerated movement, referred to as creep bursts, occur annually. During these episodes, the rock mass slides faster, with rates reaching up to $0.2$ millimetres per day. Although the origin of these bursts remains unsettled, they appear to be closely linked to water infiltration. Borehole inspections reveal that the shear zones act as preferential pathways for groundwater (see Movie S1), suggesting higher permeability than the surrounding rock. Hydrogeological surveys have identified springs within the landslide area, indicating active subsurface water flow \cite{frei2008, biorn-hansen2019, sena13groundwater}. However, pressure sensors record negligible pore pressure within the shear zones (water-head is consistently below the shear zone, orange line in Fig \ref{fig:boreholeKH0218}a), and in addition only very weak correlations are observed between fluid pressure and creep bursts (e.g., compare orange and pink lines in Figure \ref{fig:boreholeKH0218}a). Together with the observed rapid and localized fluid flow within the shear zone (see Movie S1), observations suggest water effects may not arise from elevated pore-pressure, but rather from fluid flow, indicating a gap in the current understanding and prediction of shear zone creeping dynamics.
\section*{When seepage forces control creep accelerations} \label{sec2}
\begin{figure}[!ht]
\centering
\includegraphics[width=0.95\linewidth]{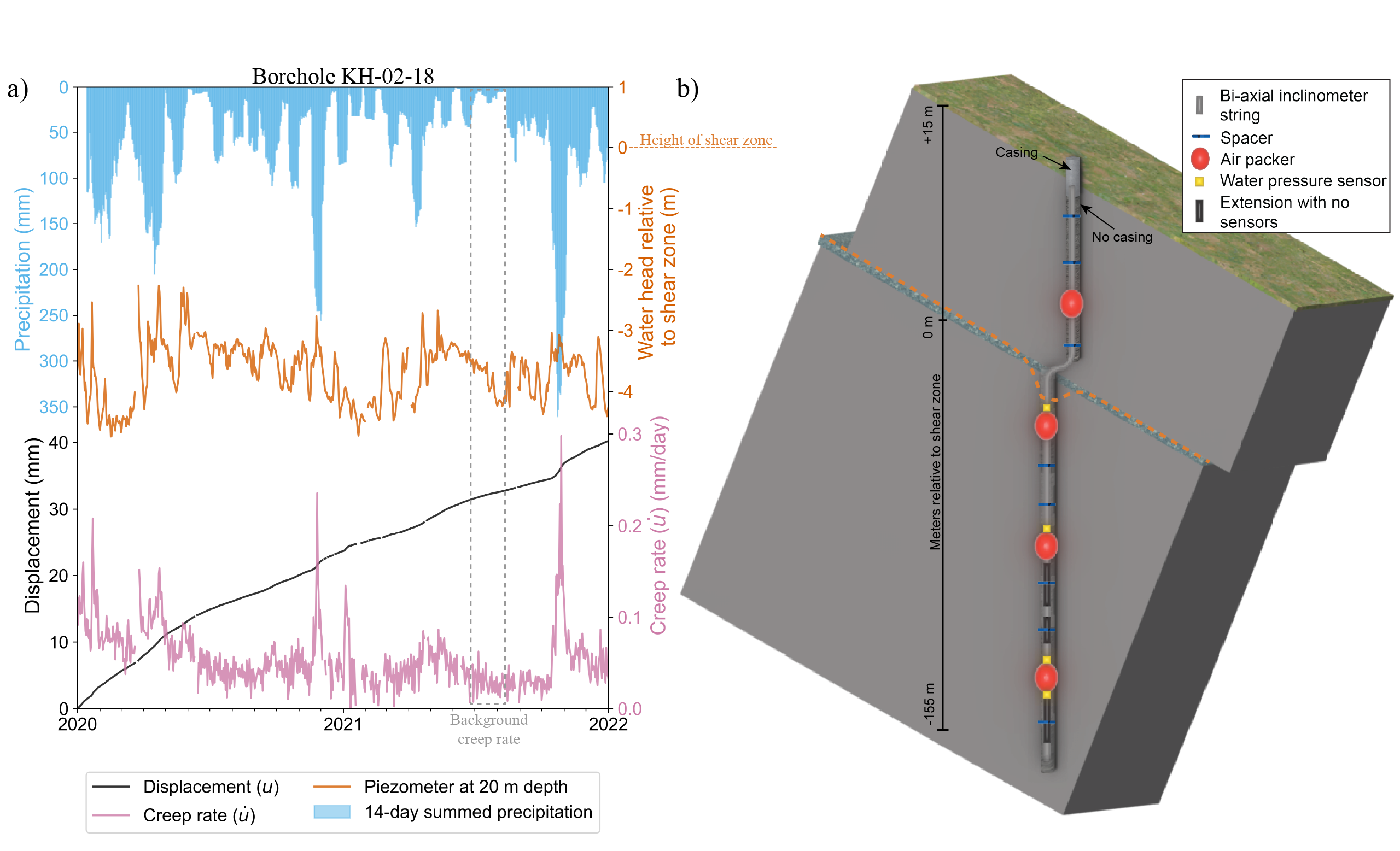}
\caption{Marginal fluid pressure variations are measured in the shear zone of \AA knes landslide. (a) Time series of different data recorded at the level of the shear zone in the borehole KH-02-18: water head relative to the depth of the shear zone (orange), cumulative creep displacement over the period of observation (black) and creep velocity (pink) featuring seasonal bursts that correlate well with the precipitation summed over the last 14 days shown in blue. The dashed grey box highlights the background creep rate that is reached after a dry period. (b) Sketch of the borehole instrumentation adapted from \cite{aspaas2024} and interpretation of the water head profile (dashed orange line). The borehole causes a draining effect, leading to local drop in the water head, which can explain the negative values shown in panel (a).}
\label{fig:boreholeKH0218}
\end{figure}
%To obtain a first-order estimate of the shear zone viscosity, we first focus on the background creep rate, which is the slowest creep rate that is reached over dry periods (see grey dashed box in Figure \ref{fig:boreholeKH0218}a). The creep viscosity of granular shear zone and fractured rock reported in the literature has order of magnitude between $10^{12}-10^{16}$ [Pa·s]. 
Next, we focus on the data collected at the borehole KH-02-18, which has the advantage of being located immediately upstream of two springs (S29 and S30 in Figure \ref{fig:setup}a) monitored in past hydrogeological surveys \cite{frei2008,biorn-hansen2019}. As detailed in Methods, they report seasonal variations of the flow rate between $v_{w,\rm bg}\sim10^{-3}$ [m/s] during the dry period and $v_{w,\rm bg}\sim10^{-1}$ [m/s] during the wet period. The hydrological fieldwork \cite{frei2008} also provides estimates of the hydraulic conductivity $K$ of the shear zone, between 0.1-1 [mm/s], which is about two orders of magnitude larger than the hydraulic conductivity measured across the wall rock ($K_{wr}\sim10^{-6}-10^{-5}$ [m/s]\cite{aspaas2024}). Together with the visual inspection of the borehole (Movie S1), these measurements confirm that the fractured and granulated shear zone is a favourable underground flow path across the landslides. Next, we use our creep equation (\ref{equ:creep_equ}) to quantitatively test the hypothesis that the force resulting from seepage across the shear zone controls seasonal creep accelerations. We first consider the background creep state, which corresponds to the slowest creep rate $\dot{u}_{\rm bg}=0.028$ [mm/day] that is reached at the end of dry periods. In the borehole data, background creep is reached between July and August 2021 (grey dashed box in Figure \ref{fig:boreholeKH0218}a). We then investigate the change in mechanical conditions causing the creep acceleration measured later that year, which coincides with important precipitation (blue area in Figure \ref{fig:boreholeKH0218}a) and culminates in a creep burst on November 2021 with $\dot{u}_{\rm peak}=0.3$ [mm/day].

From the creep equation (\ref{equ:creep_equ}), the changes in the force balance subsequent to intense rainfall can be threefold. First, variations in pore fluid pressurization through change in the $\lambda$-term. Yet, pressure measurements at the depth of the shear zone (orange curve in Figure \ref{fig:boreholeKH0218}a) recorded negligible fluid pressure with seasonal variations corresponding to less than two percent of the normal stress ($\lambda\ll1$). Even in absence of local pressure variations, creep acceleration can be triggered by a remote pocket of high fluid pressure through the elastic coupling described by the second right hand side term of Eq. (\ref{equ:creep_equ}). Such effect is well documented in the context of fluid injection along critically stressed faults, where slip front migrates faster than pressure front \cite{bhattacharya2019fluid,Viesca2021selfsimilar,saez2022threedimensional}. Yet, we can demonstrate (see Methods) that the \AA knes landslide is rather in the "marginally pressurized" regime characterized by pressure fronts migrating faster than slip front and, therefore, ignore elastic interactions at the origin of creep bursts. Consequently, the seepage forces are the only plausible triggering mechanism for this landslide, such that subtracting the creep equation (\ref{equ:creep_equ}) at peak creep rate from its value at background creep rate, leads to: 
\begin{equation}
    \dot{u}_{\rm peak} = \dot{u}_{\rm bg}+\frac{\omega}{\eta}\Big(v_{w,\rm peak}-v_{w,\rm bg}\Big).
    \label{equ:creep_burst_equ}
\end{equation}
Equation \ref{equ:creep_burst_equ} is used to estimate the viscosity of the shear zone and is evaluated using the other model parameters that are well constrained by data recorded at borehole KH-02-18 or previous field works in \AA knes (see the list and references in Table \ref{Tab:dimensionless_SI}). The estimated value $\eta\sim10^{12}-10^{13}$ [Pas] is consistent with the creep viscosity measured for clay-rich sediments (e.g., glacier till \cite{fowler1993creep}) and fractured rock \cite{talukdar2021viscoplastic} reported in the literature and brings further support to the importance of seepage forces in the seasonal variations of creep observed in \AA knes.

\begin{figure}[!ht]
\centering
\includegraphics[width=0.95\linewidth]{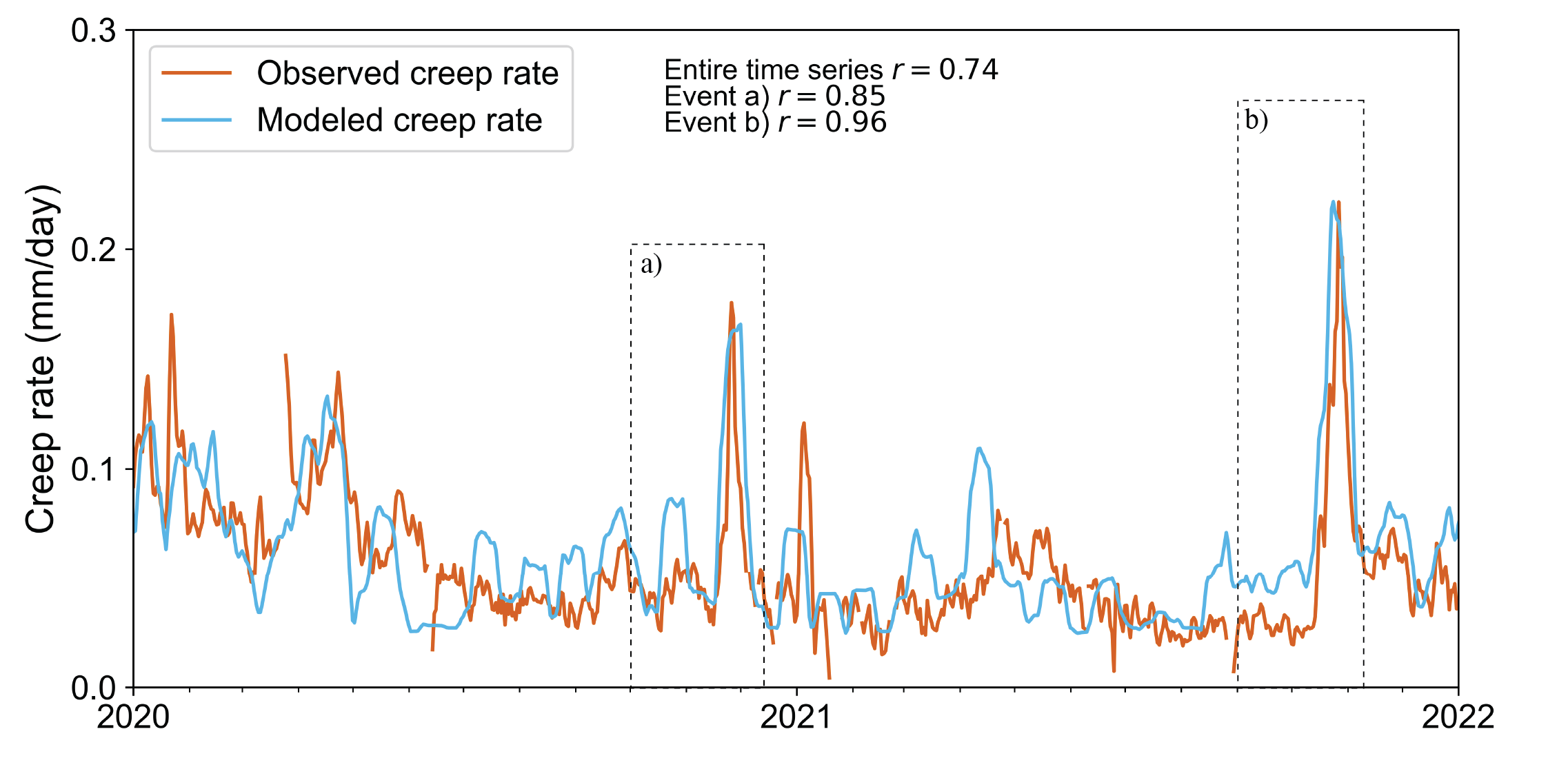}
\caption{The creep equation (\ref{equ:creep_equ}) and its parameters computed from the monitoring data allow for constructing a reliable prediction of the creep rate that is solely based on the accumulated precipitation over the last fourteen days. The model prediction (blue line) shows excellent correlation with the observed borehole KH-02-18 velocity (orange line), particularly during the creep burst events highlighted by the boxes a) and b). It represents a promising tool to decipher any changes in the mechanical behaviour of the shear zone in the future.}
\label{fig:seepage}
\end{figure}

Using the estimated viscosity and the rest of the model parameters (Table \ref{Tab:dimensionless_SI}), we further demonstrate the pivotal role of seepage force by applying the proposed creep equation (\ref{equ:creep_equ}) to predict the entire time series of creep rate recorded at the borehole KH-02-18 (Figure \ref{fig:seepage}). In absence of a continuous monitoring of the flow rate in the springs, we use the variations of the cumulated precipitation (blue area in Figure \ref{fig:boreholeKH0218}a) to interpolate the time evolution of $v_{\rm w}(t)$ between the two extrema $v_{w,\rm bg}$ and $v_{w,\rm peak}$ reported by the hydrological surveys. In hydrology, precipitation and flow rate are related through the ratio between the catchment area and the cross-sectional area. Remarkably, with this single fit parameter, our creep model successfully reproduces the entire time history of creep rates observed at borehole KH-02-18, as shown in Fig \ref{fig:seepage}.
\section*{Implications for landslide dynamics}
\begin{figure}[!ht]
\centering
\includegraphics[width=0.95\linewidth]{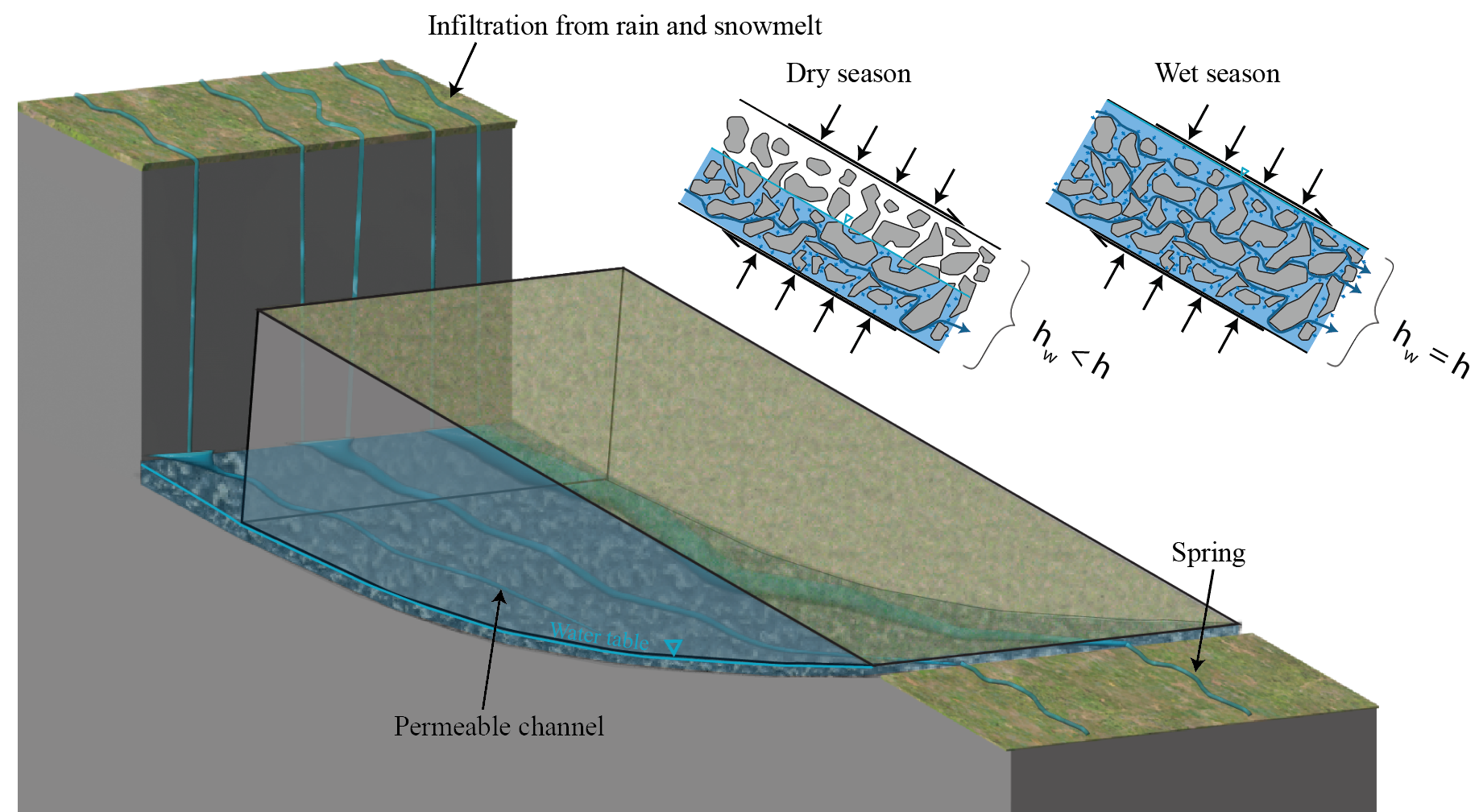}
\caption{A schematic illustration of water infiltration and its effects on subsurface water flow and water table height. Variations in hydraulic conductivity within the landslide shear zone create preferential flow paths, referred to as permeable channels. As illustrated in the comparison between dry and wet conditions, increased infiltration raises the height of water saturation $h_{\rm w}$ within the shear zone and, thereby, the average flow rate. According to the seepage force equation, higher infiltration leads to stronger seepage forces acting across the shear zone.}
\label{fig:summary_fig}
\end{figure}
The relationship between water infiltration and creep burst activities of the \AA knes landslide has long been suspected \cite{blikra2012evaluering}. The recent deployment of state-of-the-art monitoring equipment capable of measuring the history of pore pressure at the level of the shear zone reveals negligible fluid pressurisation observed throughout the year and raises the question of the origin of the destabilising effect of underground fluid flow in the landslide. In this work, we derive a generic model combining basic stress balance with granular rheology and pore fluid flow to disentangle the contributions of water flow and water pressure. Using the model, we demonstrate that seepage forces resulting from fluid flow through the porous shear zone is the dominant mechanism that promotes seasonal creep acceleration in this landslide. 

In absence of significant pressure variations in the shear zone, seasonal fluctuation of the average fluid flow rate $v_{\rm w}$ probably mainly results from change of the saturation height $h_{\rm w}$ within the shear zone between drier and wetter seasons (Figure \ref{fig:summary_fig}). In addition, fluid flow through a layer of sheared sediments and debris (e.g., in landslide \cite{tacher2005modelling} or at the base of glacier \cite{bouchayer2024multiscale}) typically forms heterogeneous structure made of channels with higher permeability and flow rate. A similar flow structure is likely expected through the shear zone of the \AA knes landslide, as evidenced by the spatial distribution of springs at the surface (Figure \ref{fig:summary_fig}). Yet, seepage forces are mildly affected by this heterogeneous structure, as comparable seepage forces, scaling as $\omega v_{\rm w}\sim v_{\rm w}/K$, are expected in permeable channels with elevated flow rate and in less permeable regions with slower flow rates.

The paradigm shift from pressurisation to seepage driven creep burst in \AA knes (and also applicable to other slides, e.g.  \cite{tang2019geohazards}) has strong implication for hazard mitigation strategies. Notably, the drainage of the shear zone should be implemented with care to avoid increasing the flow rate. The proposed creep model can be continuously updated from the precipitation measurement and included in early warning systems. Interestingly, the prediction of creep rate shown in Figure \ref{fig:seepage} is made with constant hydro-mechanical parameters of the shear zone and corresponds to a state of secondary creep, where fluctuations of the rate are only the response of fluctuations in the driving forces. Any future deviations between the model predictions and the measured creep rate may serve as a warning of   active changes of the shear zone properties, and potentially the onset of the tertiary creep regime leading to failure. In this context, the creep rate prediction can also be used as a benchmark for measuring the impact and efficiency of mitigation strategies after their implementation. 

Different geological contexts can produce  an unstable slope overriding on top of a shear zone comprising saturated debris, to which our model is directly applicable. At the Earth's surface, the rheology of these granular media plays a pivotal role in the deformation of landscapes and the formation of landslides and rockslides, as well as on the friction at the base of glaciers. The strength of the model proposed in this paper is its compact and generic formulation, relying on a minimal number of parameters. The novelty of our approach is to consider micro-mechanical fluid-solid interactions arising at the scale of the shear zone thickness, that are often overlooked in models assuming an infinitely thin friction plan. Our theoretical work, coupled with unprecedented natural data, reveal how seepage forces at the scale of the shear zone can play a pivotal in the rheology of creeping landslides and should be considered in addition to the well-established pressurisation effect. In this context, our model can be coarse-grained in the following rate-hardening frictional rheology:
\begin{equation}
    \tau(\dot{u},p,v_{\rm w}) = \tan\psi(\sigma_n-p) + \frac{\mu}{h}\dot{u} - \frac{h\rho_{\rm w}g\phi}{K}v_{\rm w},
\end{equation}
which includes both the weakening due to both fluid pressurisation and seepage forces. Two key aspects of our model, the viscosity and seepage forces, are relevant to shallow shear planes and deeper tectonic faults \cite{jerolmack_viewing_natphysrev_2019,viesca2019slow} and directly emerge from the finite thickness of the shear zone. Thus, the insights from the natural laboratory of \AA knes motivates future work to disentangle the solid and fluid interactions arising within the thickness of shear zones and governing their creeping rheology, with implications for the stability of natural slope, friction at the base of glaciers, and fluid-induced activation of faults.
\section*{Methods}

\subsection*{Seepage forces in poro-elastic media}
The system shown in Figure \ref{fig:concept} is modelled as a poroelastic medium comprising a mixture of solid rock particles, of density $\rho_r$, and pore fluid of density $\rho_{\rm w}$. The total density of the medium depends on the porosity $\phi$ and is $\rho_t = (1-\phi)\rho_r+\phi\rho_{\rm w} = \rho_s + \rho_f$. The solid $\rho_s$ and fluid $\rho_f$ density can be used to express the momentum balance of the two phases. For the solid phase, as well as in the dry wall rock above the shear zone, the momentum balance equation is expressed as function of the effective stress carried by the solid matrix, as follows:
\begin{equation}
    \nabla\cdot\mat{\sigma} + \rho_s\vec{g} + \vec{f}^{\rm seep} = 0.
    \label{equ:mom_solid}
\end{equation}
where $\vec{f}^{\rm seep}$ is the seepage force. In Eq. (\ref{equ:mom_solid}), $\mat{\sigma}$ corresponds to the effective stress carried by the solid matrix, which leads to the following tensor of elastic deformation:
\begin{equation}
    \mat{\varepsilon}^{\rm el}=\frac{1}{2G}\Big(\mat{\sigma}-\frac{\nu}{1+\nu}\mathrm{tr}(\mat{\sigma})\mat{I}\Big).
    \label{equ:elastic_deformation}
\end{equation}
Here $G$ and $\nu$ are respectively the shear modulus and Poisson's ratio of the rock. In addition to elastic strain, the total deformation of the landslide ($\mat{\varepsilon}=\mat{\varepsilon}^{\rm el}+\mat{\varepsilon}^{\rm pl}$) also includes viscoplastic deformation, which is assumed to be localised onto the shear zone (Figure \ref{fig:concept}). The only non-zero component of $\mat{\varepsilon}^{\rm pl}$ corresponds to the in-plane viscoplastic shear strain $\gamma\equiv\varepsilon^{\rm pl}_{xy}$ and is linearly proportional to the frictional shear stress in the viscous rheology of the shear plane (\ref{equ:viscous_model}).

In Eq. (\ref{equ:mom_solid}), the body force $\vec{f}^{\rm seep}$ corresponds to the seepage force exerted by the pore fluid on the solid matrix. The momentum balance for the fluid phase is expressed as follows:
\begin{equation}
    -\nabla p + \rho_f\vec{g} - \vec{f}^{\rm seep} = 0,
    \label{equ:mom_fluid}
\end{equation}
where the body force exerted by the solid matrix on fluid flow corresponds to $-\vec{f}^{\rm seep}$ from action-reaction principle (a.k.a. Newton's third law). Assuming that flow in the porous shear zone is in the Darcy regime, the vector of seepage forces corresponds to:
\begin{equation}
    \vec{f}^{\rm seep} = \frac{\mu_{\rm w}}{k}\vec{q} - \rho_{\rm w}(1 - \phi)\vec{g},
    \label{equ:darcy_seepage}
\end{equation}
such that the combination of Eqs. (\ref{equ:mom_fluid}) and (\ref{equ:darcy_seepage}) leads to Darcy's law for flow in porous media:
\begin{equation}
    \vec{q} = -\frac{k}{\mu_{\rm w}}(\nabla p-\rho_{\rm w}\vec{g}),
    \label{equ:darcy}
\end{equation}
defined from the shear zone permeability $k$, the fluid viscosity $\mu_{\rm w}$ and the fluid flow rate through the porous shear zone:
\begin{equation}
\vec{q}=\phi(\vec{v}-\dot{\vec{u}}), 
\end{equation}
which is expressed from the difference between the absolute velocities of the fluid $\vec{v}$ and solid $\dot{\vec{u}}$ phases. For the creep regime studied in this paper, we assume that $\vec{v}\gg\dot{\vec{u}}$. 

The sum of equations (\ref{equ:mom_solid}) and (\ref{equ:mom_fluid}) leads to the total momentum balance equation of the poroelastic medium:
\begin{equation}
    \nabla\cdot\mat{\sigma}^{\rm tot} + \rho_t\vec{g} = 0,
\label{equ:mom_total}
\end{equation}
and the well-established relation between the effective and the total stress tensors:
\begin{equation}
    \mat{\sigma}^{\rm tot} = \mat{\sigma} - p\mat{I}.
\end{equation}
Note that the equation above is written with the convention that compressive stresses have a negative sign, such that the effective normal stress entering the shear zone rheology (Eq. \ref{equ:viscous_model}) corresponds to $\sigma'_n=-(\sigma^{\rm tot}_{yy}+p)$. Pore fluid pressure reduces $\sigma'_n$ and, thereby, the frictional force that opposes creeping and deformation of the shear zone. In addition to this hydrostatic weakening caused by fluid pressurisation, the solid matrix is also stressed by the seepages forces of Eq. (\ref{equ:darcy_seepage}), which can be rewritten as:
\begin{equation}
    \vec{f}^{\rm seep} = -\nabla p + \rho_f\vec{g},
    \label{equ:seepage_force}
\end{equation}
which represents the hydrodynamic effects of pore fluid flow through the permeable shear zone.

\subsection*{Micromechanical origin of seepage forces}
The expression for the seepage forces given in Eq. (\ref{equ:seepage_force}) was obtained above using continuum mechanics and mixture theory. To gain further insights on the nature of seepage forces, we show next how Eq. (\ref{equ:seepage_force}) can derive from the action of pore fluid on the solid matrix at the micromechanical scale shown in the inset of Figure \ref{fig:concept}. At the grain scale, fluid-solid interaction is characterized by Cauchy law and the tractions $\vec{t}_{\rm w}$ exerted along $\Gamma$, the surface with normal vector $\vec{n}$ marking the boundary between the two phases:
\begin{equation}
    \vec{t}_{w} = \mat{\sigma}_{\rm w}\cdot\vec{n}.
    \label{equ:tractions}
\end{equation}
From Newton's third law, $\mat{\sigma}_{\rm w}$ corresponds to the stress tensor either in the solid phase or the fluid phase. For the incompressible fluid phase, the stress tensor writes:
\begin{equation}
    \mat{\sigma}_{\rm w} = -p\mat{I} + \mat{\tau}_{\rm w}x,
\end{equation}
and consists of the sum of pressure and viscous shear stresses $\mat{\tau}_{\rm w}$. The fluid-solid interaction can then be divided into two contributions: a normal traction ($p\mat{I}\cdot\vec{n}$) resulting from fluid pressure and a tangential contribution ($\mat{\tau}_{\rm w}\cdot\vec{n}$) resulting from viscous shear stress. The vector of seepage forces $\vec{f}^{\rm seep}$ is the body force resulting from the coarse graining of these fluid-solid tractions:
\begin{equation}
    \vec{f}^{\rm seep} = \frac{1}{V}\int_{\Gamma} \vec{t}_{w}\d\Gamma = \frac{1}{V}\int_{\Gamma}\Big( -p\mat{I} + \mat{\tau}_{\rm w}\Big)\cdot\vec{n}\d\Gamma.
\end{equation}
In the equation above, the representative elementary volume $V$ consists of the sum of the volumes of fluid $V_f$ and solid $V_s$ phases in the shear zone and $\Gamma$ is the surface marking the boundary between them. The force per unit volume due to normal traction is equivalent to buoyancy force and can be integrated using the divergence theorem and recalling the definition of the porosity ($V_s=(1-\phi)V$):
\begin{equation}
    \vec{f}^{n} = -\frac{1}{V}\int_{\Gamma} p\mat{I}\cdot\vec{n} \;\d \Gamma = -\frac{1}{V}\int_{V_s} \nabla \cdot (p\mat{I}) \;\d \Omega = -\frac{V_s}{V}\nabla p = -(1-\phi)\nabla p.
    \label{equ:seepage_normal}
\end{equation}
The tangential traction can be estimated by idealizing pore fluid flow as a collection of pipes cutting through the porous matrix. Along the pipes, the tangential traction exerted on the wall derives from the Hagen-Poiseuille flow profile \cite{Feder2022physics}, such that the force per unit volume due to tangential traction can be written as a line integral along the pipes axis: 
\begin{equation}
\vec{f}^{t} = \frac{1}{V}\int_{\Gamma} \mat{\tau}_{\rm w}\cdot\vec{n} \;\d \Gamma = -\frac{1}{V}\int_{L_p}\int_0^{2\pi}\frac{R}{2}(\nabla p-\rho_{\rm w}\vec{g}) \;R\d \theta\d s,
\end{equation}
with $R$ corresponding to the radius of the pipes. The expression above can be simplified further using the definition of the volume $V_f=\phi V$ and density $\rho_f=\phi\rho_{\rm w}$ of the fluid phase:
\begin{equation}
     \vec{f}^{t} = -(\nabla p-\rho_{\rm w}\vec{g})\frac{1}{V}\int_{L}\int_0^{2\pi}\frac{R^2}{2} \;\d \theta\d s = -(\nabla p-\rho_{\rm w}\vec{g})\frac{V_f}{V} = -\phi\nabla p + \rho_f\vec{g}.
     \label{equ:seepage_tangential}
\end{equation}
The seepage forces correspond to the resultant vector from the sum of the normal (Eq. \ref{equ:seepage_normal}) and tangential (Eq. \ref{equ:seepage_tangential}) fluid-solid interactions and is identical to the one obtained from macroscopic stress balance (Eq. \ref{equ:seepage_force}).

\subsection*{A depth-average model}
The balance of stress on the solid matrix, Eq. (\ref{equ:mom_solid}), in the $x$-direction is:
\begin{equation}
    \frac{\partial \sigma_{xx}}{\partial x} + \frac{\partial\sigma_{xy}}{\partial y} + \frac{\partial\sigma_{xz}}{\partial z} + \rho_s g \sin\theta + f^{\rm seep}_{x} = 0.
    \label{equ:x-momentum}
\end{equation}
Next, we aim at expressing the divergence of stress in terms of the longitudinal slip displacement $u_x$ using the elastic deformation of the sliding block (Eq. \ref{equ:elastic_deformation}). First, the elastic constitutive relationship (Eq. \ref{equ:elastic_deformation}) is used to express stress in the sliding block. Assuming that the displacement in the transverse direction is negligible, the third term on the left hand side of Eq. (\ref{equ:x-momentum}) writes:
\begin{equation}
    \frac{\partial\sigma_{xz}}{\partial z} = 2G\frac{\partial \varepsilon^{\rm el}_{xz}}{\partial z} = G\frac{\partial^2 u_x}{\partial z^2}.
\end{equation}
Similarly, the first term of the momentum balance (Eq. \ref{equ:x-momentum}) can be expressed as:
\begin{equation}
    \frac{\partial \sigma_{xx}}{\partial x} = \frac{2G}{1-\nu}\frac{\partial \varepsilon^{\rm el}_{xx}}{\partial x} + \frac{\nu}{1-\nu}\frac{\partial \sigma_{yy}}{\partial x} = \frac{2G}{1-\nu}\frac{\partial^2 u_x}{\partial x^2} + \frac{\nu}{1-\nu}\frac{\partial \sigma_{yy}}{\partial x}.
    \label{equ:sigma_xx}
\end{equation}
Next, the thin-strip geometry of the sliding block (visible from the cross section in Figure \ref{fig:setup}b) enables two useful simplifications. First, the longitudinal variations of the shear and effective normal stresses can be neglected in front of their transverse variations:
\begin{equation}
\frac{\partial}{\partial y}\Big(\sigma_{yy},\sigma_{xy}\Big)\sim\rho_rg\gg\frac{\partial}{\partial x}\Big(\sigma_{yy},\sigma_{xy}\Big)\sim\rho_r g\frac{\partial H}{\partial x}. 
\end{equation}
As a result, the normal stress variation can be dropped in the right hand side of Eq. (\ref{equ:sigma_xx}), such that the gradient of the effective longitudinal stress simplifies to:
\begin{equation}
    \frac{\partial \sigma_{xx}}{\partial x} = \frac{2G}{1-\nu}\frac{\partial^2 u_x}{\partial x^2}.
    \label{equ:sigma_xx_simplified}
\end{equation}
Second, the stress balance can be solved in average over $H$ as the depth of the shear zone is much smaller than the other dimensions of the landslide:
    \begin{equation}
        \frac{1}{H}\int_0^H\Big(\frac{2G}{1-\nu}\frac{\partial^2 u_x}{\partial x^2}+ \frac{\partial\sigma_{xy}}{\partial y} + G\frac{\partial^2 u_x}{\partial z^2} + \rho_s g \sin\theta\Big) \d y - \frac{1}{H}\int_0^{h_{\rm w}} f^{\rm seep}_{x}\d y = 0.
        \label{equ:mometum_integration}
    \end{equation}
The height of the water-saturated zone $h_{\rm w}$ can be different than the thickness of the shear zone and can fluctuate with rain input and seasonality (Figure \ref{fig:summary_fig}). The first integral of Eq. (\ref{equ:mometum_integration}) can be rewritten as:
\begin{equation}
G\Big(\frac{2}{1-\nu}\frac{\partial^2 u}{\partial x^2}+\frac{\partial^2 u}{\partial z^2}\Big) + \frac{\sigma_{xy}|_{y=H}-\sigma_{xy}|_{y=0}}{H}+ \rho_rg\sin\theta\Big(1 -\phi\frac{h}{H}+\phi\frac{\rho_{\rm w}}{\rho_r}\frac{h_{\rm w}}{H}\Big),
\label{equ:gap_int_solid}
\end{equation}
with free surface and frictional stress boundary conditions respectively at the top ($\sigma_{xy}|_{y=0}=0$) and at the base ($\sigma_{xy}|_{y=H}=\tau_f$) of the sliding block and $u$ that is defined as the average displacement across the thickness of the landslide:
\begin{equation}
    u \equiv\frac{1}{H}\int_0^H u_x\d y.
    \label{equ:average_displ}
\end{equation}
Similarly, the $x$-component of the flow rate can be averaged across the thickness of the porous shear zone
\begin{equation}
    v_{\rm w} \equiv\frac{1}{h}\int_0^h v_x\d y,
\end{equation}
and used to integrate the second term on the left hand side of Eq. (\ref{equ:mometum_integration}) together with Darcy's law (Eq. \ref{equ:darcy}):
\begin{equation}
\frac{1}{H}\int^{h_{\rm w}}_0 \Big( \phi v_x\frac{\mu_{\rm w}}{k} - \rho_{\rm w}g\sin\theta(1 - \phi)\Big)\d y = \frac{\phi\rho_{\rm w}g h}{KH}v_{\rm w} - \rho_{\rm w}g\sin\theta (1 - \phi)\frac{h_{\rm w}}{H}.
\label{equ:gap_int_fluid}
\end{equation}
The smallness of the following term:
\begin{equation}
    \phi\frac{h}{H}+(1-2\phi)\frac{\rho_{\rm w}}{\rho_r}\frac{h_{\rm w}}{H}\ll1,
\end{equation}
is used to simplify the different gravitational contributions resulting from the combination of Eqs. (\ref{equ:gap_int_solid}) and (\ref{equ:gap_int_fluid}). Dividing by $\rho_r g\cos\theta$, the balance of stress writes:
\begin{equation}
    \frac{GH}{\sigma_n}\Big(2(1-\nu)^{-1}\frac{\partial^2 u}{\partial x^2}+\frac{\partial^2 u}{\partial z^2}\Big) -\frac{\tau_f}{\rho_r gH\cos\theta} + \tan\theta+ \frac{h\rho_{\rm w}\phi}{KH\rho_r \cos\theta}v_{\rm w} = 0.
    \label{equ:stress_balance_final}
\end{equation}
Similarly, the normal stress at the base of the landslide can be approximated as $\sigma_n=\rho_r gH\cos\theta$. Last, stress balance (\ref{equ:stress_balance_final}) is combined to the constitutive rheology of the creeping (Eq. \ref{equ:viscous_model}):
\begin{equation}
    \frac{\tau_f}{\sigma_n} = \tan\psi(1-\lambda) + \mu\frac{\dot{u}}{h\sigma_n},
\end{equation}
to express the creep equation (Eq. \ref{equ:creep_equ}) used in the manuscript
\begin{equation}
    \dot{u}\frac{\mu}{h\sigma_n}=\tan\theta(1-\alpha+\lambda\alpha) + \frac{GH}{\sigma_n}\Big(2(1-\nu)^{-1}\frac{\partial^2 u}{\partial x^2}+\frac{\partial^2 u}{\partial z^2}\Big) +  v_{\rm w}\frac{h\rho_{\rm w}\phi }{KH\rho_r\cos\theta}.
    \label{equ:creep-rate-full}
\end{equation}
For writing Eq. (\ref{equ:creep-rate-full}) as Eq. (\ref{equ:creep_equ}), we define the viscous coefficient
\begin{equation}
    \eta \equiv \frac{\mu}{h\sigma_n},
    \label{equ:viscous_coeff}
\end{equation}
the seepage coefficient
\begin{equation}
    \omega \equiv \frac{h\rho_{\rm w}\phi }{KH\rho_r\cos\theta},
    \label{equ:seepage_coeff}
\end{equation}
and the elastic coefficient
\begin{equation}
    \kappa \equiv \frac{GH}{\sigma_n}.
    \label{equ:elastic_coeff}
\end{equation}
\section*{Creep burst equation}
The creep equation (Eq. \ref{equ:creep-rate-full}) governs the migration of slip front over the landslide and has the form of a diffusion equation within the shear plane $y=0$ that results from the coupling of elasticity and viscous friction \cite{viesca2019slow}:
\begin{equation}
    \dot{u} = S_u + \nabla\cdot(\mat{D_u} \nabla u),
    \label{equ:diffusion_creep}
\end{equation}
with $S_u$ being the source term that corresponds to the local driving force. Neglecting the slight anisotropy due to Poisson's effect, the visco-elastic diffusivity $\mat{D_u}$ can be described by the following scalar:
\begin{equation}
    D_u = \frac{2GhH}{\mu},
\end{equation}
which has magnitude $D_u\sim10^{-3}-10^{-2}$ [m$^2$/s] using the parameters listed in Table \ref{Tab:dimensionless_SI}. Over the typical rise time of creep bursts ($t_{\rm cb}\sim$ one day to one week), elastic stress redistribution across the landslide arises over diffusion length scale ($\sqrt{D_ut_{\rm cb}}\sim10-80$ [m]), much smaller than the dimension of the landslide. Moreover, the hydraulic transport within the permeable shear plane is also described by a diffusion equation resulting from Darcy's law and fluid mass balance:
\begin{equation}
    \dot{H}_{\rm w} = S_H + \nabla\cdot(D_H\nabla  H_{\rm w}).
    \label{equ:diffusion_fluid}
\end{equation}
In Eqn.(\ref{equ:diffusion_creep}), $H_{\rm w}=y+p/(\rho g)$ corresponds to the hydraulic head, $S_H$ is the hydraulic source term and $D_H$ is the hydraulic diffusivity defined as function of water compressibility $c=4\times10^{-10}$ [Pa$^{-1}$]:
\begin{equation}
    D_H = \frac{k}{\mu_w c\phi} = \frac{K}{\rho_{\rm w} g c \phi} \sim 10^{2}-10^{3} [\mathrm{m}^2\mathrm{/s}].
\end{equation}
Moreover, the ratio of the visco-elastic and hydraulic diffusivities can be used to probe the competition between the migration of slip front and pressure front:
\begin{equation}
    \frac{D_u}{D_H}=2c(1-\nu)^{-1}G\frac{H}{h}\frac{\omega}{\eta}.
\end{equation}
Across the shear zone of \AA knes landslide, pressure front migrates much faster than slip front ($D_u/D_H\sim10^{-5}$), such that the response of the system is similar to that of \textit{marginally pressurised fault} in the context of fluid-induced seismicity \cite{bhattacharya2019fluid,Viesca2021selfsimilar,saez2022threedimensional} and explains the negligible contribution of non-local elastic interactions at the origin of creep bursts.
\section*{Hydrogeological surveys}
In \AA knes, the permeable shear zone corresponds to a preferential flow path that forms the spring horizon shown in Figure \ref{fig:setup}a where it outcrops at the surface. Within this spring horizon, springs S29 and S30 are located immediately downstream from borehole KH-02-18. In August 2008, \cite{frei2008} measured flow rates across the landslide by tracking the propagation of chemical tracers over the A-A' transect (Figure \ref{fig:setup}b) and estimated flow velocities between one and five millimetres per second. This measurements arose over a dry month (August) and we assume that it is representative of the minimal background flow rate in the shear zone ($v_{w,\rm bg}=10^{-3}$ [m/s]). An independent hydrogeological campaign measured variations of the volumetric flow rate out of the middle spring horizon at different periods of the years 2018-2019. They reported more than two orders of magnitude difference between the minimum ($0.02$ litre per second in August 2019) and the maximum volumetric flow rate ($5.9$ litre per second in April 2019). As a result, we assume that the background flow rate is two orders of magnitude smaller than the peak flow rate ($v_{w,\rm peak}=10^{-1}$ [m/s]).

In addition, \cite{frei2008} quantified the permeability along the flow path by assuming that fluid flow between the injection point and the sampling point arises through an effective fracture with constant width $2b$ (see Eq. 4.10 of \cite{frei2008}):
\begin{equation}
    2b = \sqrt\frac{12\nu_{\rm w}\tau_f^2X^2}{gt_a\Delta H_{\rm w}}\in[0.05,0.18]\:\mathrm{mm}.
    \label{equ:effective_frac}
\end{equation}
using the nominal values for the kinematic viscosity of water $\nu_{\rm w}=1.3\cdot10^{-6}$ [m$^2$/s], tortuosity factor $\tau_f=1.5$. In the equation above, $t_a$ is the average flow time, with $X$ and $\Delta H_{\rm w}$ being respectively the horizontal distance and change in hydraulic head between injection and sampling point. Assuming that fluid transport between the injection and sampling points arises as Darcy flow through a porous media, Equation (\ref{equ:effective_frac}) can be rewritten in term of the hydraulic conductivity:
\begin{equation}
    K = \frac{\phi X^2}{t_a\Delta H} = (2b)^2 \frac{g\phi}{12\nu_{\rm w}\tau_f^2}\in[0.2,2]\:\mathrm{mm/s}.
\end{equation}

\begin{table}
\everymath{\displaystyle}
\renewcommand{\arraystretch}{2.8}
     \begin{tabular}{| l | c | c | c | l } \hline   
        Physical parameters & Variables & Values for \AA knes & References \\ \hline
        Depth of the shear zone  & $H$ & 15 [m] & \cite{aspaas2024}\\ \hline
        Thickness of the shear zone  & $h$ & 0.1 [m] & \cite{langeland2019a}\\ \hline
        Hydraulic conductivity & $K$ & $10^{-4}$ [m/s] & \cite{frei2008} \\ \hline
        Shear zone viscosity & $\mu$ & $10^{13}$ [Pa s] & Post-processed using Eq. (\ref{equ:creep_burst_equ}) \\ \hline
        Shear zone porosity & $\phi$ & 30\% & {\tiny Nominal value for uncons. debris} \\ \hline
        Slope angle & $\theta$ & 30\textdegree & \cite{ganerod2008geological} \\ \hline
        Angle of repose & $\psi$ & 25\textdegree-29\textdegree & \cite{langeland2014utvikling,groneng2009shear} \\ \hline
        Rock mass density & $\rho_r$ & $2900$ [kg/m$^3$] & Nominal value for Gneiss \\ \hline
        Poisson's ratio & $\nu$ & 0.3 & Nominal value for Gneiss \\ \hline
        Elastic shear modulus & $G$ & $8\times10^{9}$ [Pa] & Nominal value for Gneiss\\ \hline
    %    Average yearly creep rate & $\langle \dot{u} \rangle$ & 6 [cm/year] & \cite{aspaas2024} \\ \hline
        Shear stress & $\tau=\rho_r gH\sin\theta$ & 0.21 [MPa] & Using parameters above\\ \hline
        Normal stress & $\sigma_n=\rho_r gH\cos\theta$ & 0.37 [MPa] & Using parameters above \\ \hline
        Viscous coefficient & $\eta=\frac{\mu}{h\sigma_n}$ & $3\times10^{8}$ [s/m] & Using parameters above \\ \hline
        Seepage coefficient & $\omega=\frac{h}{H} \frac{\rho_{\rm w} \phi}{K \rho_r \cos\theta}$ & $8$ [s/m] & Using parameters above \\ \hline
        Elastic coefficient & $\kappa=\frac{GH}{\sigma_n}$ & $3\times10^{5}$ [m] & Using parameters above \\ \hline
        Factor of safety & $\alpha=\frac{\tan\psi}{\tan\theta}$ & $0.85$ & Using parameters above \\ \hline
    \end{tabular}
   \caption{List of the physical parameters used to analyse creep burst at the borehole KH-02-18 in \AA knes landslide.}
   \label{Tab:dimensionless_SI}
\end{table}
\section*{Supplemental material}
Movie S1: Video acquired by an inspection camera inserted in the borehole KH-02-18 on (26.10.2019) before the installation of the monitoring equipment. The depth from surface is shown on the bottom left corner. Once the camera reached the depth of the shear zone (at about fifteen meters), it crossed a region of damaged and granulated rock mass, which reveals evidence of underground fluid flowing through this permeable damaged material.

\section*{Data availability}
The data and scripts used to produce the figures of this publication will be archived on Zenodo and made accessible at the time of publication.

\section*{Acknowledgments}

This project received funding from the Research Council of Norway (grant agreement No. 345008 UNLOC), from the Norwegian Water Resources and Energy Directorate (NVE) and from the European Research Council under the European Union's Horizon 2020 research and innovation program (grant agreement No. 101019628 BREAK). 
\bibliography{main.bbl}% common bib file
\end{document}